\documentstyle[preprint,aps]{revtex}

\tighten

\begin{document} \draft

\title{Inelastic light, neutron, and X-ray scatterings related to the
heterogeneous elasticity of glasses}
\author{E. Duval, L. Saviot}
\address{LPCML, Universit\'{e} Lyon I, UMR-CNRS 5620  43 boulevard du 11
Novembre, 69622 Villeurbanne Cedex, France}
\author{A. Mermet}
\address{ESRF, B.P. 220, 38043 Grenoble, France}
\author{L. David}
\address{GEMPPM, UMR-CNRS 5510, INSA-Lyon, 69621 Villeurbanne Cedex, France}
\author{S. Etienne}
\address{LPM, UMR-CNRS 7556, Ecole des Mines, 54042 Nancy Cedex, France}
\author{V. Bershtein}
\address{Ioffe Physico-technical Institute, Russian Academy of Sciences,
St. Petersburg 194021, Russia}
\author{A. J. Dianoux}
 \address{ILL, B.P. 156, 38042 Grenoble, France}

\date{\today}
\maketitle

\begin{abstract}
The effects of plasticization of poly(methyl methacrylate) glass on the
boson peaks observed by  Raman and neutron scattering are
compared. In plasticized glass the cohesion heterogeneities are responsible
for the neutron boson peak and partially for the Raman one, which is
enhanced by the composition heterogeneities. Because  the composition
heterogeneities have a size similar to that of the cohesion ones and form
quasiperiodic clusters, as observed by  small angle X-ray scattering, it is
inferred that the cohesion heterogeneities in a normal glass form
nearly periodic arrangements too. Such structure at the nanometric scale
explains the linear dispersion of the vibrational frequency versus the
transfer momentum observed by inelastic X-ray scattering.
\end{abstract}
\newpage

\section{INTRODUCTION}

The excess of low-frequency vibrational density of states (VDOS) in
comparison with the Debye regime, called boson peak, is quasi-universal in
glasses. It is generally admitted that its origin lies in nanometric
fluctuations, its frequency being related to the correlation
length. S. Elliott \cite{Ell92} interpreted the boson peak by density
fluctuations which scatter and consequently localize the acoustic phonons,
within a continuous random network, or a homogeneous structure (all points
in the glass are considered as equivalent). More recently several authors
interpreted the boson peak by atomic force or elastic constants
fluctuations, still in homogeneous \cite{Schi98,Tar01} structures.

As these interpretations show it, the excess of VDOS can be ascribed to
fluctuations of density or of elastic constants within a homogeneous
structure. In
spite of that, several interpretations have postulated a heterogeneous glassy
nanostructure to account for the fluctuations: in this case the correlation
length corresponds to the heterogeneity size
\cite{Phil81,Duv90,Mal91,Ach91}. However, up to now, no clear density
nanoheterogneities have been observed by either small angle X-ray or neutron
scattering, nor by electron microscopy. It is why the existence of cohesion
heterogeneities was suggested \cite{Duv90}. Within this latter view, it is
assumed that more cohesive nanodomains are separated by softer zones
\cite{Ach95,Mer96}.

Several arguments, also based on the concept of heterogeneities though
transient as they pertain to the liquid state, come as a strong support for
the existence of cohesion heterogeneities within the glassy state.
According to
Stillinger, "the vitrifying liquid is viewed as a dynamic patchwork of
relatively strongly bonded (but amorphous) molecular domains that are
separated by irregular walls of weakened bonds" \cite{Stil88}. Such a
description was confirmed, on the one hand by the multidimensional nuclear
magnetic resonance technique, that made evident the existence of transient
heterogeneities in the supercooled state \cite{Tra98}, and on the  other
hand by simulation works that showed the existence of domains inside which
atoms
are more mobile \cite{Kob97}. Stillinger added "the mean domain diameter rises
as temperature declines to minimize wall free energy, but the process is
self-limiting due to intradomain frustration energy" \cite{Stil88}.
Theories related to frustration, developped by Kivelson and Tarjus
\cite{Kiv95,Tarj00} do also predict a supermolecular heterogeneity termed
as ``frustration limited domains''. From all these considerations, it
appears natural that the so-called dynamical heterogeneities freeze upon
cooling below the glass transition temperature, under the form of an
inhomogeneous elastic constant network. An experimental evidence of the
heterogeneous glass structure
is found in a very recent paper \cite{Vid00}, in which  the heterogeneous
molecular cooperative relaxation, at a temperature slightly lower than the
glass transition, is experimentally demonsrated, by using a nanometric
electric probe.

Very recently, numerical simulations using a Lennard-Jones
potential, were able to confirm the existence of heterogeneities of elastic
constants in disordered matter. Wittmer et al.
\cite{Wit01} find that, in a 2-dimensional amorphous system, the atomic
displacements become nonaffine with respect to the sample deformation
below a length scale of 30 to 50 molecular sizes; these displacements are
found to be correlated in
regions of a size similar to this length scale. On the other hand, Viliani
et al. \cite{Vil01}, by determining the atomic potential energy variation
induced by the normal vibrational modes, show that  the correlation between
``softest''  atoms is at least as strong as in the whole sample. Such a
heterogeneous elasticity is not far from the glass description by a
heterogenous distribution of positive and negative internal stresses as
suggested by Alexander \cite{Alex98} and simulated by Kustanovich and Olami
\cite{Kus00}.

In the first part of this paper, by comparing the boson peaks observed by
respectively  Raman  and inelastic neutron scattering of a plasticized
polymeric glass, the relation between the boson peak and an experimentally
observed heterogeneous
nanostructure will be established. In the
second part, it will be emphasized that the linear dispersion of the
vibrational frequency $\nu$ versus the momentum transfer $Q$, that is
generally observed by inelastic X-ray scattering \cite{Mas96,Ben96,Mer98},
does not necessarily prove the vibrational energy propagation and is in
agreement with
the heterogeneous glass cohesion at the nanometric scale \cite{Duv98}.

\section{RAMAN AND NEUTRON BOSON PEAKS OF PLASTICIZED POLY(METHYL
METHACRYLATE)}

Recently the low-frequency Raman scattering (LFRS) of glassy poly(methyl
methacrylate) (PMMA) plasticized by 23 (mass)$\%$ of dibutyl phtalate (DBP)
was studied. The plasticized PMMA will be symbolized by PMMA/DBP. It was
observed an enhancement of the Raman boson peak compared with the one that
is expected for PMMA/DBP (Figure-1)  \cite{Duva98}. On the contrary, no
enhancement of the VDOS in the spectral range of the boson peak was
observed by inelastic neutron scattering (Figure-2). One can observe in
Figure-2 that the VDOS calculated by addition of the PMMA and DBP VDOS in
PMMA/DBP fits perfectly the experimental VDOS of the plasticized PMMA
\cite{Sav01}. As far as the static structure is concerned, a correlation
peak was detected  with PMMA/DBP by small angle
X-ray scattering (SAXS) at an exchanged momentum $Q\simeq 0.15 \, \AA^{-1}$
\cite{Duva98}. Such a correlation peak is absent in pure PMMA.
This momentum $Q$ value corresponds to a period of about $40\, \AA$ in a
quasiperiodic network. The obtained period is close to that deduced from
the position of the boson peak in PMMA, using a model of inhomogeneous
nanostructure \cite{Merm96}.  It is remarkable that similar observations
were done with another polymeric glass: an enhancement of the boson peak of
glassy poly(vinyl chloride) was observed by Raman scattering \cite{Duv95},
and not by neutron scattering \cite{Zor98},  after plasticization by $10\%$
of dioctyl phtalate. This meaningful difference between the Raman and neutron
boson peak behaviours can be explained by the different origins of  the
Raman and neutron scattering excesses, as explained further below.

The observation of a correlation peak in PMMA/DBP demonstrates that the
distribution of the plasticizer in PMMA is inhomogeneous: regions poor in
DBP are separated by DBP-rich zones, and are arranged quasiperiodically.
The neutron boson peak corresponds to the VDOS excess. From experiment, this
inhomogeneous and quasiperiodic distribution of DBP in PMMA has no effect
on either the position or the intensity of the VDOS excess, which is
approximately
determined by the sum of the VDOS of both components in PMMA/DBP
\cite{Sav01}. Several informations can be deduced from this result. (1) As
the boson peak is dependent on the intermolecular bonding, it is deduced
that macromolecules of PMMA interact principally with other PMMA
macromolecules. (2) The frequency of the boson peak depends on the
correlation length- or heterogeneity size. Consequently, if the
plasticization does not change the boson peak frequency, the correlation
lengths in PMMA and DBP in PMMA/DBP are approximately the same as in pure
PMMA and pure DBP.

As it is well known, the Raman boson peak intensity is proportional to the VDOS
excess and to the frequency-dependent light-vibration coupling tensor
$C_{\alpha\beta}(\nu)$. Therefore, the enhancement of the Raman boson peak
(Figure-2) comes from $C_{\alpha\beta}(\nu)$, which is expressed as follows
for an isotropic material \cite{Jac81,Duv93}:

\begin{eqnarray}
\label{cret1}
C_{\alpha\beta}(\nu) \propto \int\int d{\bf r_{1}} d{\bf r_{2}}
<\delta\chi_{\alpha\beta}({\bf r_{1}},\nu) \delta\chi_{\alpha\beta}({\bf
r_{2}},\nu)>
\end{eqnarray}

\noindent
$\alpha$ and $\beta$ denote the polarizations of the exciting and
scattered light respectively, $\delta\chi$ is the susceptibility
fluctuation. Brackets  denote the space correlation.
$\delta\chi_{\alpha\beta}({\bf r},\nu)$ is proportional to the strain
tensor $s_{\gamma\delta}({\bf r},\nu)$ corresponding to the vibrational
mode of frequency $\nu$ \cite{Jac81,Nov95}:

\begin{eqnarray}
\label{cret2}
\delta\chi_{\alpha\beta}({\bf r},\nu)=P_{\alpha\beta\gamma\delta}({\bf
r})s_{\gamma\delta}({\bf r},\nu)
\end{eqnarray}

\noindent
$P_{\alpha\beta\gamma\delta}({\bf r})$ is the elasto-optic tensor. The
coupling coefficient $C_{\alpha\beta}(\nu)$ has two origins \cite{Jac81}:
the elastic or mechanical fluctuations, which partially localize the
vibrational modes, and the fluctuations of
$P_{\alpha\beta\gamma\delta}({\bf r})$:

\begin{eqnarray}
\label{cret3}
P_{\alpha\beta\gamma\delta}({\bf r})=P_{\alpha\beta\gamma\delta}+\delta
P_{\alpha\beta\gamma\delta}({\bf r})
\end{eqnarray}

The observation by SAXS of a correlation peak at a transfer momentum Q
testifies the existence of periodic composition- or density-fluctuations
\cite{Duva98}. Surely, these fluctuations induce periodic
elasto-optic fluctuations
$\delta P_{\alpha\beta\gamma\delta}(r)$ such that:

\begin{eqnarray}
\label{cret4}
\delta P_{\alpha\beta\gamma\delta}(r)=\delta
P^{0}_{\alpha\beta\gamma\delta} cos(\frac{2\pi a_{\ell}}{r})+ harmonic
\,\,terms
\end{eqnarray}

\noindent
with $a_{\ell}$ the period of PMMA/DBP observed by SAXS
($a_{\ell}=\frac{2\pi}{Q}$). The strain tensor in (\ref{cret2}) is
proportional to the space phase term
$\exp(i{\bf k}.{\bf r})$ of the corresponding acoustic vibrational mode.
${\bf k}$ is the wavevector that characterizes the  (non-necessarily
propagating) mode. It is such that $k\simeq 2\pi\nu/v$, $v$ being the sound
velocity. From Equations \ref{cret1} and \ref{cret2}, the contribution  of
the elasto-optic fluctuations to the light-vibration coupling coefficient
will be maximum for $k=2\pi/a_{\ell}$, i.e., for:

\begin{eqnarray}
\label{cret5}
\nu\simeq \frac{v}{a_{\ell}}
\end{eqnarray}

In a pure non-plasticized PMMA glass no density fluctuations with a size
larger than 0.5 nm are observed,
so that it is expected that the mechanical fluctuations have a stronger
contribution to $C(\nu)$ than the electric ones. Now, it was shown that the
mechanical contribution to $C(\nu)$ presents no maximum and has a power-law
frequency dependence \cite{Duv93,Nov95}. Consequently, it was deduced that
the Raman boson peak frequency is related to the neutron boson peak one, i.e.,
to the frequency of the VDOS excess. From different models
\cite{Ell92,Duv90}:

\begin{eqnarray}
\label{cret6}
\nu\simeq \frac{v}{d_{\ell}}
\end{eqnarray}

\noindent
$d_{\ell}$ being the correlation length or the heterogeneity size of
glass. As the enhanced Raman boson peak of PMMA/DBP appears at a frequency
close to that of pure PMMA (Figure-1), it is deduced  that
$a_{\ell}\simeq d_{\ell}$ and that the correlation of the elastic
fluctuations in PMMA is not very different from that of the composition
fluctuations in PMMA/DBP. In some way, the plasticizer ``decorates'' the
elastic constant fluctuations. It is tempting to conclude that there are
nearly periodic elastic constant arrangements in glasses, and that the boson
peak corresponds to the Van Hove singularities of these periodic
arrangements. As it will be emphasized in the following part, this explains
the linear dispersion $\nu(Q)$ observed by inelastic X-ray scattering.

\section{INELASTIC X-RAY SCATTERING AND HETEROGENEOUS NANOSTRUCTURE OF
GLASSES. DISCUSSION}

The linear dispersion of longitudinal acoustic excitations $\nu(Q)$,
observed by inelastic X-ray
scattering (IXS) in many glasses  has been a subject of controversy. While
some authors
\cite{Mas96,Ben96,Mer98} observed the linear dispersion up to frequencies
higher than that of the boson peak $\nu_{b}$, others claimed that the
dispersion linearity stops at a frequency lower than $\nu_{b}$, at least in
the case of silica glass \cite{For96}. It is indeed surprising to observe
propagating modes, at frequencies higher than $\nu_{b}$, since
acoustic phonon attenuation measurements in silica \cite{Zhu91} and
PMMA \cite{Mor96} showed that propagation ceases at frequencies lower than
$\nu_{b}$. Recently it was shown that the linear dispersion observed in the
region of boson peak can be explained by non-propagating modes in an
elastically heterogeneous structure \cite{Duv98}. This interpretation is
now simply described in relation to the previous part.

The static structure factor of glasses, as observed by SAXS, shows no feature
indicating the presence of nanoheterogeneities with a size larger than 0.5
nm. As emphasized above, this
non-obervation is in agreement with the absence of contrasted density
nanoheterogeneities but not in disagreement with the presence of elastic
constant ones. As already suggested, the plasticizer highlights the
elastic constant nanoheterogeneities like it does in semicrystalline
polymers \cite{Bal87}. The broad correlation peak, that is observed by SAXS
with PMMA/DBP \cite{Duva98}, indicates (1) that at the nanometric scale there
is a distribution of periods and (2) that the periodicities are extended on
relatively short distances. What is observed by the IXS of a glass? In the
spectral range of the boson peak, IXS simply probes the dynamic structure
factor of the
vibrating nanometric structure. In the network submitted to a vibration,
whose wavelength
is close to the heterogneity size, the elastic periodicity is
transformed into a density periodicity so that a dynamic correlation peak is
observed. From (\ref{cret6}) and $a_{\ell}\simeq d_{\ell}$, the dispersion
relation $\omega\simeq vQ$ ($\omega=2\pi\nu$) is obtained. The linearity
can be observed in the whole spectral range of the boson peak because of
the above mentionned period- or size heterogeneity- distribution. By scanning
in frequency $\nu$ or in transfer momentum Q, one scans the size
distribution.

Following this interpretation, the observed IXS peaks correspond to the
modes at
the edge of the Brillouin zones or at the Van Hove singularities of
nearly periodic arrangements comprising several elastic constant
heterogeneities. Consequently, the observed linear dispersion in the
spectral range of the boson peak does not correspond to the dispersion
curve in a single periodic or continuous network. Furthermore, these modes
observed by IXS do not propagate phonons (group velocity equal to zero at
the edge of Brillouin zone), in agreement with other measurements
\cite{Zhu91,Mor96}. Finally, it is noted that this interpretation is not in
contradiction, on the one hand, with  the topological origin of the
$Q^{2}$-dependence of the vibration attenuation \cite{Mas96,Ben96,Grig01}
and, on
the other hand, with numerical simulations which seem to confirm the linear
dispersion and the $Q^{2}$-dependence of the attenuation \cite{Ang00,Mos01}.

\section{CONCLUSION}

The boson peaks observed by Raman and neutron scatterings and the Brillouin
peak observed by inelastic X-ray scattering in the $nm^{-1}$ transfer momentum
range are in agreement with the nanometric cohesion heterogeneity of
glasses. The enhancement of the Raman boson peak in plasticized PMMA
compared to that in pure PMMA and the non-enhancement of the neutron boson
peak suggest that the plasticizer highlights the cohesion heterogeneities
by density or composition ones and, consequently, by elasto-optic constant
heterogeneities. On the other hand, these observations and the  relatively
broad correlation peak observed by small angle X-ray scattering in
plasticized PMMA show that the density heterogeneities, and consequently
the cohesion ones in pure glass, form periodic arrangements. Such a topological
nanostructure possibly explains the linear dispersion observed by inelastic
scattering: the peaks, that are observed in the spectral range of the boson
peak, can correspond to the Van Hove singularities of the nearly periodic
arrangements.

\newpage

\begin{figure}
\label{f1}
\caption{Reduced Raman spectrum of PMMA/DBP at T=140 K (full circles),
compared to the addition of the spectra of pure PMMA and DBP at the same
temperature (open circles). The inset shows the reduced Raman spectrum of
DBP at 140 K.}
\end{figure}

\begin{figure}
\label{f2}
\caption{Vibrational density of states divided by the square of frequency,
$G(\nu)/\nu^{2}$, deduced from the inelastic neutron scattering at $30\,K$.
PMMA: empty circles; PMMA/DBP, experiment: full squares; calculated: empty
squares; DBP: full circles. The
normalization is obtained by an arbitrary equalization of $G(\nu)/\nu^{2}$
at 1 THz.}
\end{figure}

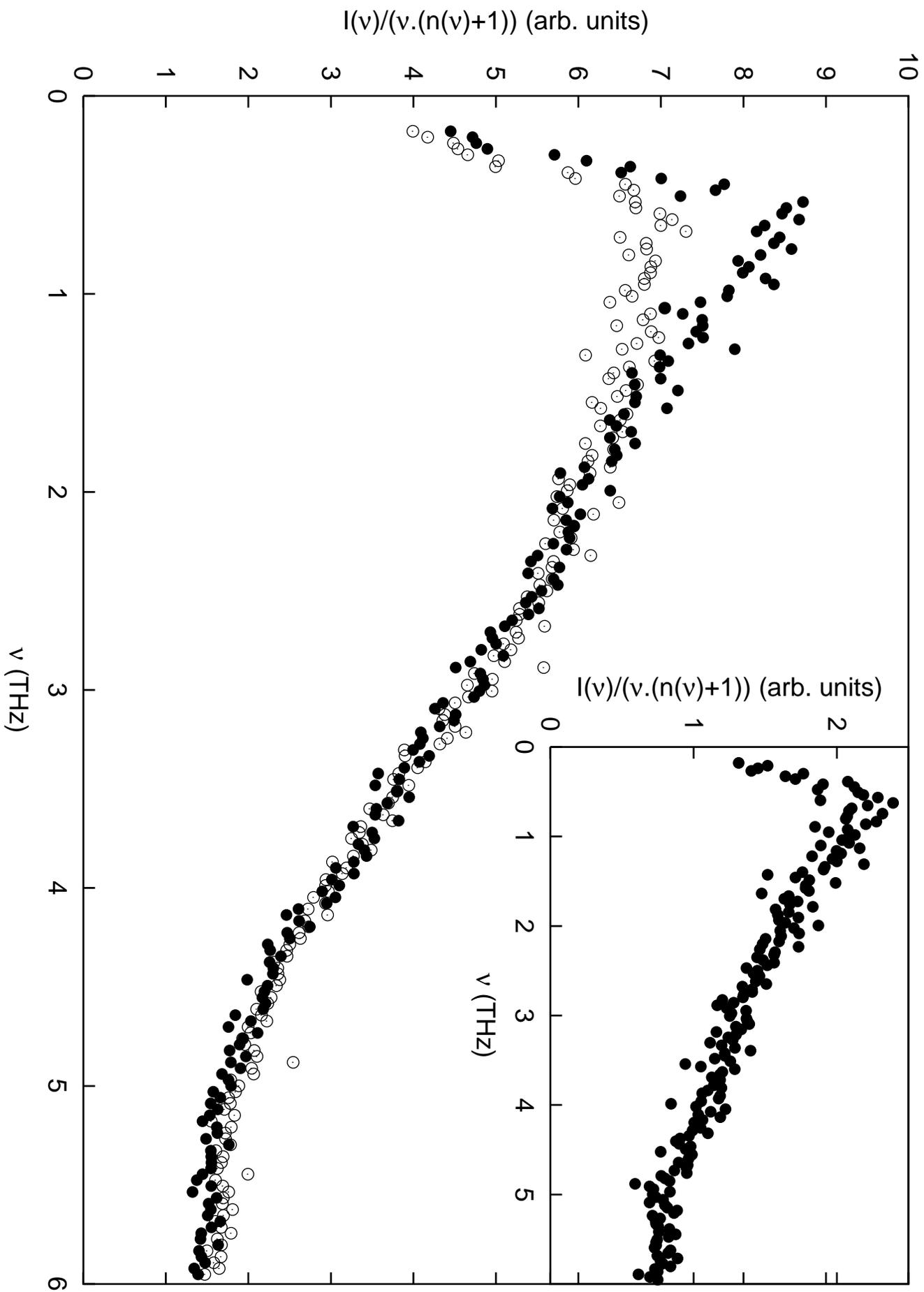
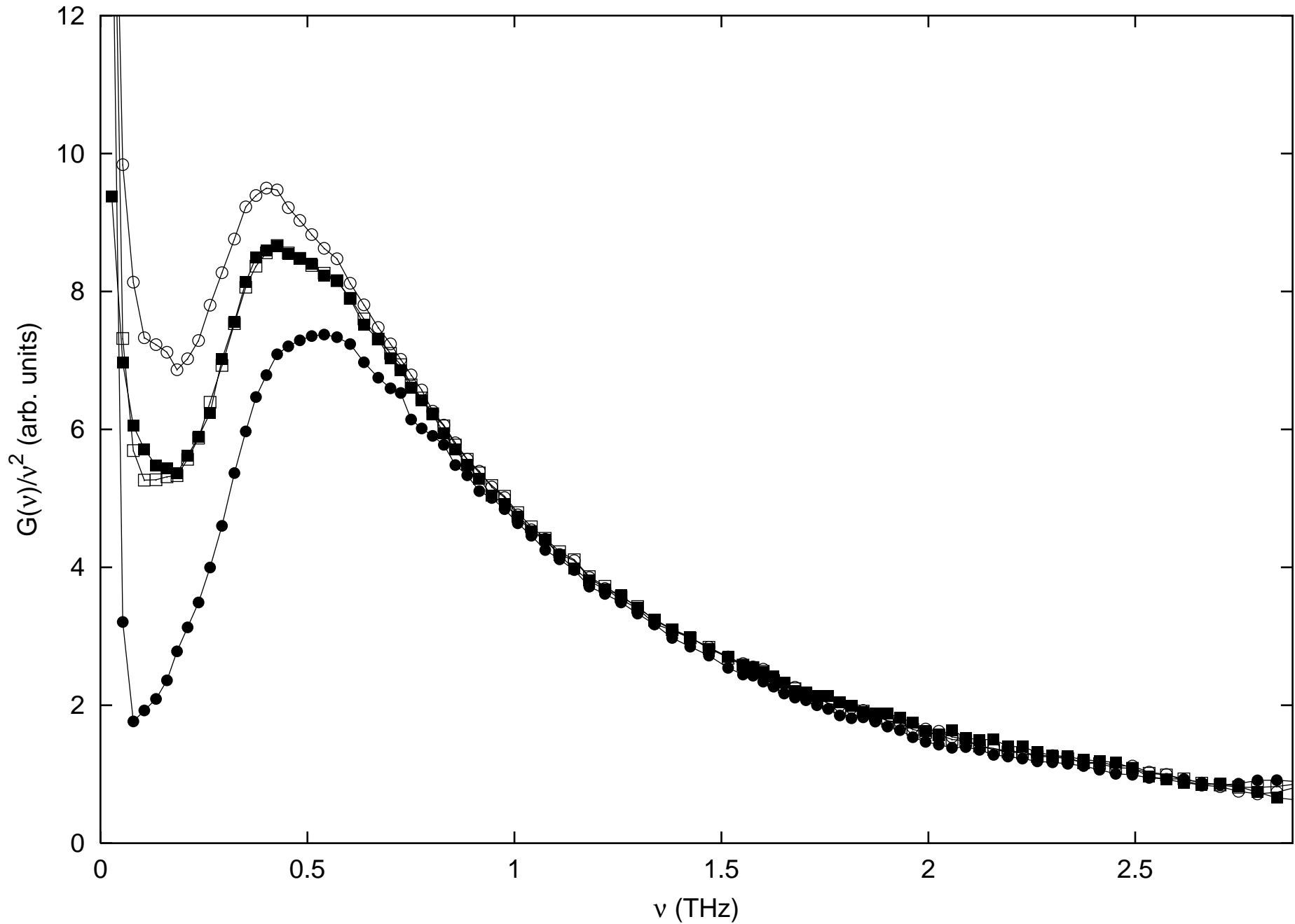
\end{document}